\documentclass[12pt]{iopart}
\usepackage[utf8]{inputenc}
\usepackage{latexsym}
\usepackage{graphicx}
\usepackage{epsfig}
\usepackage{amsopn}
\usepackage{amsfonts}
\usepackage{amssymb}
\usepackage{braket}
\usepackage{citesort}
\usepackage{epstopdf}
\usepackage{braket}
\newcommand{\Op}[1]{\ensuremath{\mathsf{\hat{#1}}}}
\newcommand{\dd}{\ensuremath{\, \textnormal{d}}} 

\newcommand{\tgt}{\operatorname{tgt}}
\renewcommand{\Re}{\mathfrak{Re}}
\renewcommand{\Im}{\mathfrak{Im}}

\begin{document}
\title{Controlling the transport of an ion: Classical and quantum
  mechanical solutions} 

\author{H A F\"urst$^1$, M H Goerz$^2$, U G Poschinger$^1$, M
  Murphy$^3$, S Montangero$^3$, T
  Calarco$^3$, F Schmidt-Kaler$^1$, K Singer$^1$, C P Koch$^2$}

\address{$^1$QUANTUM, Institut f\"ur Physik, Universit\"at Mainz,
  D-55128 Mainz, Germany}
\address{$^2$Theoretische Physik, Universit\"at Kassel,
  Heinrich-Plett-Stra{\ss}e 40,  D-34132 Kassel, Germany}
\address{$^3$Institut f\"ur Quanteninformationsverarbeitung,
  Universit\"at Ulm, D-89081 Ulm, Germany}
\ead{christiane.koch@uni-kassel.de}
\begin{abstract}
We investigate the performance of different control techniques for
ion transport
in state-of-the-art segmented miniaturized ion
traps. We employ numerical optimization of classical
trajectories and quantum wavepacket propagation as well as analytical
solutions derived from invariant based inverse engineering and
geometric optimal control. We find that accurate shuttling can be
performed with operation times below the
trap oscillation period. The maximum speed is limited
by the maximum acceleration that can be exerted on the ion.
When using controls obtained from classical dynamics for wavepacket
propagation, wavepacket squeezing is the only quantum effect that
comes into play for a large range of trapping parameters. We show that
this can be corrected by a compensating 
force derived from invariant based inverse engineering, without a
significant increase in the operation time. 
\end{abstract}


\maketitle

\section{Introduction}

Trapped laser-cooled ions represent a versatile experimental platform
offering near-perfect control and tomography of a few body system in
the classical and quantum
domain~\cite{CIRAC1995,Blatt2008,WINELAND1998,CASANOVA2012}. The fact
that both internal (qubit) and external (normal modes of oscillation)
degrees of freedom can be manipulated in the quantum regime allows for
many applications in the fields of quantum information processing and
quantum simulation~\cite{MONZ2011,GERRITSMA2011,RICHERME2013}.
Currently, a significant research effort is devoted to scaling
these experiments up to larger numbers of qubits. A promising
technology to achieve this goal are \textit{microstructured segmented
  ion traps}, where small ion groups are stored in local potentials
and ions are shuttled within the trap by applying suitable voltage
ramps to the trap electrodes~\cite{KIELPINSKY2002}. In order to enable
scalable experiments in the quantum domain, these shuttling operations
have to be performed such that the required time is much shorter than
the timescales of the relevant decoherence processes. At the same
time, one needs to avoid excitation of the ion's motion after the
shuttling operation. These opposing requirements clearly call for the
application of advanced control techniques.

Adiabatic ion shuttling operations in a segmented trap have been demonstrated
in Ref.~\cite{ROWE}. Recent experiments have achieved
non adiabatic shuttling of single ions within a few trap oscillation
cycles while retaining the quantum ground state of
motion~\cite{WALTHER2012,Bowler2012}. This was made possible by
finding `sweet spots'\ in the shuttling time or removal of the excess
energy accumulated during the shuttling by kicks of the trap
potential. Given the experimental constraints, it is natural to ask
what the speed limitations for the shuttling process are. The impact
of quantum effects 
for fast shuttling operations, i.e., distortions of the wavepacket,
also need to be analyzed, and it needs to be assessed whether
quantum control techniques~\cite{SomloiCP93,ZhuJCP98,ReichJCP12}
may be applied to avoid these.
Moreover, from a control-theoretical perspective and in view of
possible future application in experiment, it is of interest to
analyze how optimized voltage ramps can be obtained.
Optimal control theory (OCT) combined with
classical equations of motion was employed in Ref.~\cite{Schulz2006}
to obtain optimized voltage
ramps. Quantum effects were predicted not to play a role unless the
shuttling takes place on a timescale of a single oscillation period.
In Refs.~\cite{CHEN2011,Torrontegui2011}, control techniques such as
inverse engineering were applied to atomic shuttling problems. The
transport of atomic wavepackets in optical dipole potentials was
investigated using OCT with quantum mechanical equations of
motion~\cite{CALARCO2004,deChiaraPRA08,MurphyPRA09}.

The purpose of the present paper is to assess available
optimization strategies for the specific problem of transporting a
single ion in a microchip ion trap and to utilize them to study the
quantum speed limit for this
process~\cite{GiovannettiPRA03,CanevaPRL09}, i.e., to determine the
shortest possible time for the transport. Although parameters of the
trap architecture of Ref.~\cite{SCHULZ2008} are used
throughout the entire manuscript, we strongly emphasize that the
qualitative results we obtain hold over a wide parameter regime. They
are thus generally valid for current segmented ion traps, implemented
with surface electrode geometry~\cite{SCHULZ2008,AMINI2011}
or more traditional multilayer geometry.

The paper is organized as
follows. We start by outlining the theoretical framework in
Sec.~\ref{subs:ttcp}. In
particular we review the combination of numerical optimization with
classical dynamics in Sec.~\ref{subs:kloct} and with wavepacket motion
in Sec.~\ref{subs:qmoct}. Analytical solutions to the control problem,
obtained from the harmonic approximation of the trapping potential,
are presented in Secs.~\ref{subsec:geometric} and~\ref{subs:invmeth0}.
Section~\ref{subs:apcontrol} is devoted to the presentation and
discussion of our results. The control solutions for purely classical
dynamics of the ion, obtained both numerically and analytically, yield
a minimum transport duration as shown in Sec.~\ref{subs:kloct2}. We
discuss  in  Sec.~\ref{subs:qmprop2}, how far these solutions
correspond to the quantum speed limit. Our results obtained by
invariant-based inverse engineering are presented in
Sec.~\ref{subs:invmeth}, and we analyze the feasibility of quantum optimal
control in Sec.~\ref{subs:qmoct2}. Section~\ref{sec:concl} concludes
our paper.


\section{Methods for trajectory control and wavepacket
  propagation}\label{subs:ttcp}

In the following we present the numerical methods we employ to control
the transport of a single trapped ion. Besides numerical optimization
describing the motion of the ion either with classical mechanics or
via wavepacket propagation, we also utilize two analytical
methods. This is made possible by the trap geometry which leads to an
almost perfectly harmonic trapping potential for the ion at all
times.

\subsection{Prerequisites}
\begin{figure}[tbp]
 \centering
 \includegraphics[width=0.40\textwidth]{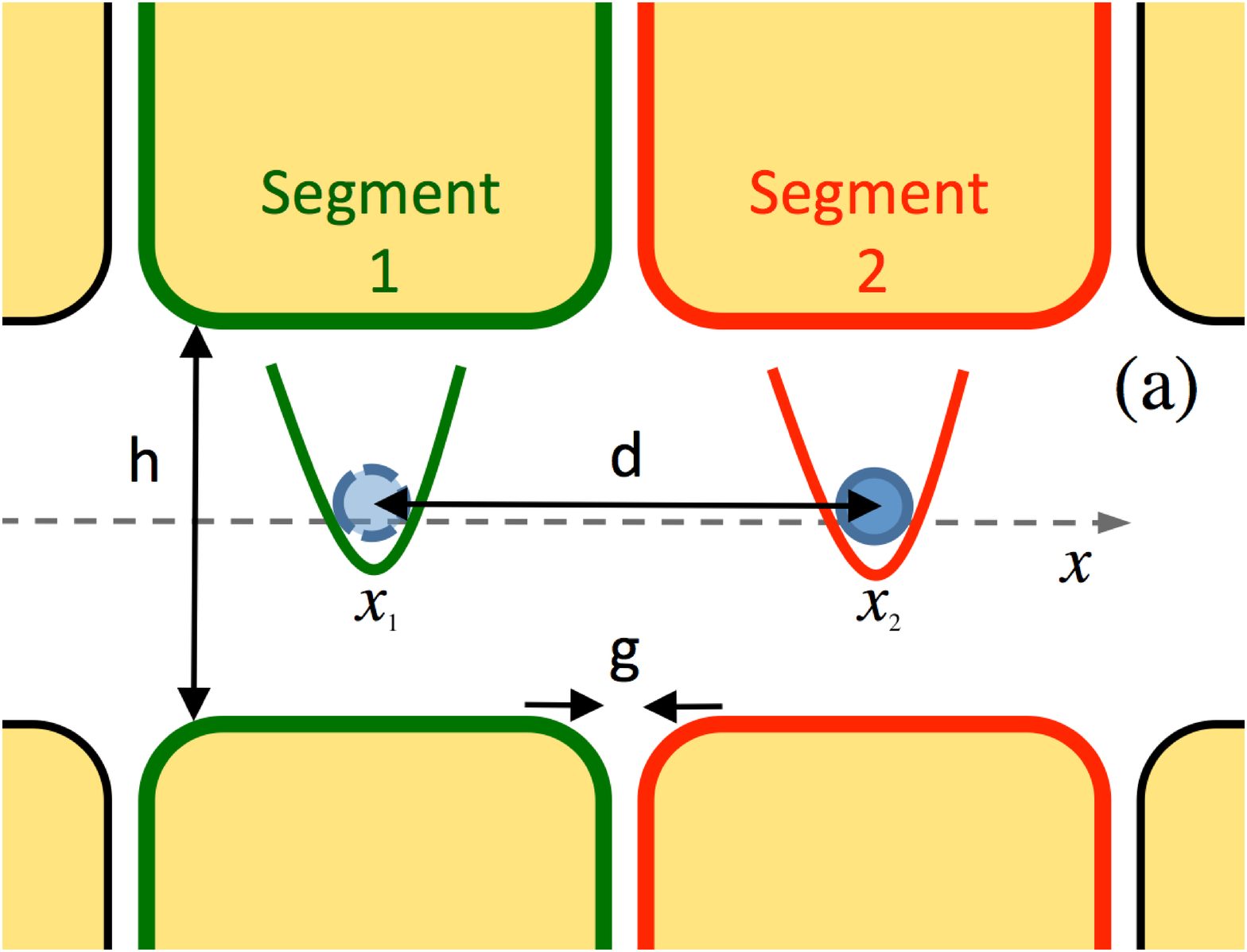}
 \hspace{0.5cm}
 \includegraphics[width=0.46\textwidth]{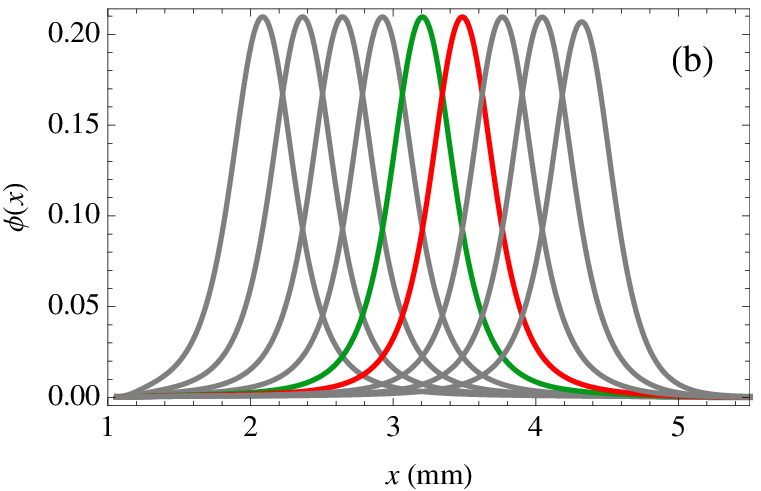}
 \caption{(a) Ion shuttling in a segmented linear trap. The
   dc electrodes form the axial potential for the ion transport along
   the $x$-axis. The rf electrodes
   for confinement of the ions along the $x$-axis are not shown. (b)
   Axial electrode potentials formed by applying a
   dc voltage to a facing pair of trap segments.
   For the specific scenario presented in this manuscript, we use $d =
   280\,\mu$m, $g=30\,\mu$m and $h=500\,\mu$m.
   Each potential is
   generated from a single pair of segments, depicted in red in
   (a) and biased to 1$\,$V with all the other dc electrodes grounded.
 }
\label{fig:electrodes}
\end{figure}

We assume ponderomotive confinement of the ion
at the rf-node of a linear segmented Paul trap and a purely
electrostatic confinement along the trap axis $x$, see Fig.~\ref{fig:electrodes}.
This enables us to treat the dynamics only along this dimension. We consider
transport of a single ion with mass $m$ between two neighboring electrodes,
which give rise to individual potentials centered at $x_1$ and $x_2$. This may
be scaled up to $N$ electrodes and longer transports without any loss of
generality.
The ion motion is controlled by a time-dependent electrostatic potential,
\begin{equation}\label{eq:V}
V(x,t)=  U_1(t)\phi_1(x)+U_2(t)\phi_2(x)\,,
\end{equation}
with segment voltages $U_i(t)$, and normal electrode potentials on the
trap axis, $\phi_i(x)$. They are dimensionless electrostatic potentials obtained
with a bias of $+1$~V at electrode $i$ and the remaining electrodes
grounded (see Fig.~\ref{fig:electrodes}(b)). 
These potentials are calculated by using a fast multipole boundary element
method~\cite{SINGER2010} for the trap geometry used in recent
experiments~\cite{WALTHER2012} and shown in Fig.~\ref{fig:electrodes}. In order to
speed up numerics and obtain smooth derivatives, we calculate values for
$\phi_i(x)$ on a mesh and fit rational functions to the resulting data. The spatial derivatives $\phi_i'(x)$ and $\phi_i''(x)$ are obtained by differentiation of the fit functions.
Previous experiments have shown that the calculated potentials allow for
the prediction of ion positions and trap frequencies with an accuracy
of one per cent~\cite{Huber2010,brownnutt2012spatially} which
indicates the precision of the microtrap fabrication process. An
increase in the precision can be achieved by calibrating the trapping
potentials using
resolved sideband spectroscopy. This is sufficient to warrant the
application of control techniques as studied here. 
For the geometry of the trap described in Ref.~\cite{WALTHER2012}, we
obtain harmonic trap
frequencies of about $\omega=2\pi\cdot$1.3~MHz with a bias voltage of -7~V
at a single trapping segment. The individual segments are spaced 280~$\mu$m
apart. Our goal is to shuttle a single ion along this distance within a time span on
the order of the oscillation period by changing the voltages $U_1$ and $U_2$,
which are supposed to stay within a predetermined range that is set by
experimental constraints. We seek to minimize the amount of motional excitation
due to the shuttling process.

\subsection{Numerical optimization with classical dynamics}
\label{subs:kloct}
Assuming the ion dynamics to be well described classically,
we optimize the time dependent voltages in order to reduce the amount
of transferred energy. This corresponds to minimizing
the functional $J$,
\begin{equation}\label{eq:J}
J = (E(T)-E_{\rm T})^2 + \sum_i \int_0^T \frac{\lambda_a}{S(t)} \Delta U_i(t)^2 \dd t\,,
\end{equation}
i.e., to miniziming
the difference between desired energy $E_{\rm T}$ and the energy
$E(T)$ obtained at the final time $T$.
$\Delta U_i(t)= U_i^{n+1}(t) - U_i^{n}(t)$ is the update of each
voltage ramp in an iteration step $n$, and the second term in
Eq.~\eref{eq:J} limits the
overall change in the integrated voltages during one iteration.
The weight $\lambda_a$  is used to tune the convergence and limit the
updates. To suppress updates near $t=0$ and $t=T$ the shape function
$S(t) \geq 0$ is chosen to be zero at these points in time. For a
predominantly harmonic axial confinement, the final energy is given by
\begin{equation}\label{eq:ET}
  E(T) = \frac{1}{2} m \dot{x}^2(T) + \frac{1}{2} m \omega^2 (x(T)-x_2)^2\,.
\end{equation}
In order to obtain transport without motional excitation, we choose
$E_T = 0$. Evaluation of Eq.~\eref{eq:ET} requires solution of
the classical equation of motion. It reads
\begin{equation}\label{eq:class:eom}
\ddot x(t) = -\frac{1}{m} \left.\frac{\partial}{\partial x}
  V(x,t)\right|_{x=x(t)} = -\frac{1}{m} \sum_{i=1}^2
U_i(t)\phi_i'\left(x(t)\right)
\end{equation}
for a single ion trapped in the potential of Eq.~\eref{eq:V} and is
solved numerically using a \textit{Dormand-Prince Runge-Kutta}
integrator~\cite{SINGER2010}.
Employing Krotov's method for optimal control~\cite{Konnov99}
together with the classical equation of motion,
Eq.~\eref{eq:class:eom}, we obtain the
following iterative update rule:
\begin{equation}\label{eqn:krotklupd}
  \Delta U_i(t) = -
  \frac{S(t)}{\lambda_{a}} p_2^{(n)}(t) \phi_i'(x^{(n+1)}(t))\,,
\end{equation}
where $n$ denotes the previous iteration step. $\mathbf{p}=(p_1,p_2)$
is a costate vector which evolves according to
\begin{equation}
  \dot{\mathbf{p}}(t) = - \left(\begin{array}{c} \frac{p_2}{m} V''(U_i(t), x(t)) \\ p_1 \end{array}\right)\,,
\end{equation}
with its `initial' condition defined at the final time $T$:
\begin{equation}\label{eqn:pTkrot}
\mathbf{p}(T) = - 2 m \left(E(T)-E_{\rm T}\right) \left(\begin{array}{c} \omega^2 (x(T) - x_2) \\ \dot{x}(T) \end{array}\right)\,.
\end{equation}
The algorithm works by propagating $x(t)$ forward in time, solving
Eq.~\eref{eq:class:eom} with an initial guess for $U_i(t)$ and
iterating the following steps until the desired value of $J$ is
achieved:
\begin{enumerate}
  \item Obtain $p(T)$ according to Eq.~\eref{eqn:pTkrot} and propagate $p(t)$ backwards in time using Eq.~\eref{eqn:pTkrot}.
  \item Update the voltages according to Eq.~\eref{eqn:krotklupd} at each time
    step while propagating $x(t)$ forward in time with the immediately
    updated voltages.
\end{enumerate}
The optimization algorithm shows rapid convergence and brings the final
excitation energy $E(T)$ as close to zero as desired.
An example of an optimized voltage ramp is shown in
Fig.~\ref{fig:guesscfoct}(a). The voltages obtained are not symmetric under
time reversal in contrast to the initial guess. This is rationalized
by the voltage updates occurring only during forward propagation which
breaks the time  reversal symmetry.
We find this behavior to be typical for the Krotov algorithm combined
with the classical equation of motion.

\subsection{Numerical optimization of wavepacket propagation}\label{subs:qmoct}

When quantum effects are expected to influence the transport, the ion has to
be described by a wave function $\Psi(x,t)$. The control target is
then to perfectly transfer the initial wavefunction, typically the
ground state of the trapping potential centered around position $x_1$,
to a target wavefunction, i.e., the
ground state of the trapping potential centered around position $x_2$.
This is achieved  by minimizing the functional
\begin{equation}\label{eq:qmkrotj}
  J = 1 - \left\vert \int\limits_{-\infty}^{\infty}
            \Psi(x,T)^* \Psi^{\tgt}(x) \dd x
         \right\vert^2
       + \int\limits_{0}^{T} \frac{\lambda_a}{S(t)} \sum_i
               \Delta U_i(t)^2 \dd t\,.
\end{equation}
Here, $\Psi(x,T)$ denotes the wave function of the single ion  propagated
with the set of voltages $U_i(t)$, and $\Psi^{\tgt}(x)$ is the
target wave function.
The voltage updates $\Delta U_i(t)$, scaling factor $\lambda_a$ and shape
function $S(t)$ have identical meanings as in
Sec.~\ref{subs:kloct}. $\Psi(x,T)$ is obtained by solving
the time-dependent Schr\"odinger equation (TDSE),
\begin{eqnarray}
  i \hbar \frac{\partial}{\partial t} \Psi(x,t) = \Op H(t) \Psi(x,t)
  = \left( -\frac{\hbar^2}{2m}\frac{d^2}{dx^2}
    + \sum_{i=1}^{N} U_i(t) \phi_i(x)
  \right) \Psi(x,t)\,.
    \label{eq:tdse}
\end{eqnarray}
As in the classical case, optimization of the transport problem is
tackled using Krotov's
method~\cite{SomloiCP93,ReichJCP12}. The update equation derived from
Eq.~\eref{eq:qmkrotj} is given by
 \begin{equation}\label{eqn:krotovupdate}
   \Delta U_i(t)
     = \frac{S(t)}{\lambda_a}
     \Im\int\limits_{x_{\min}}^{x_{\max}}
    \chi^{n}(x,t)^*\,
    \phi_i(x) \,
    \Psi^{n+1}(x,t)
    \dd x\,,
\end{equation}
with $n$ denoting the iteration step. $\chi(x,t)$ is a costate wave
function obeying the TDSE with `initial' condition
\begin{equation}\label{eqn:chioft}
\chi(x,T) =
\left[
\int\limits_{x_{\min}}^{x_{\max}}
(\Psi(x,T))^* \Psi^{\tgt}(x) \dd x \;
\right]
\Psi^{\tgt}(x,T)\,.
\end{equation}
Optimized voltages $U_i(t)$ are obtained similarly to
Sec.~\ref{subs:kloct}, i.e., one starts with the ground state,
propagates $\Psi(x,t)$ forward in time according to
Eq.~\eref{eq:tdse}, using an
initial guess for the voltage ramps, and iterates the following steps
until the desired value of $J$ is achieved:
\begin{enumerate}
\item Compute the costate wave function at the final time $T$
  according to Eq.~\eref{eqn:chioft} and
  propagate $\chi(x,t)$ backwards in time, storing $\chi(x,t)$ at each
  timestep.
\item Update the control voltages according to
  Eq.~\eref{eqn:krotovupdate} using the stored $\chi(x,t)$, while
  propagating $\Psi(x,t)$ forward using the immediately updated
  control voltages.
\end{enumerate}
Equations~\eref{eqn:krotovupdate}
and~\eref{eqn:chioft} imply a sufficiently large initial overlap
between the wave function, which is
forward propagated under the initial guess, and the target state
in order to obtain a reasonable voltage update. This emphasizes the
need for good initial guess ramps and illustrates the difficulty
of the control problem when large phase space volumes need to be
covered.

To solve the TDSE numerically, we use the Chebshev
propagator~\cite{Tal-EzerJCP84} in
conjunction with a Fourier grid~\cite{RonnieReview88,RonnieReview94}
for efficient and accurate application of the kinetic
energy part of the Hamiltonian.
Denoting the transport time by $T$ and the inter-electrode spacing by
$d$, the average momentum during the shuttling is given by $\bar{p}=m
d / T$. Typical values of these parameters yield a phase space volume of
$d \cdot \bar{p}/h\approx 10^7$.
This requires the numerical integration to be extremely stable. In
order to ease the numerical treatment, we can exploit the fact that
the wavefunction's spatial extent is much smaller than $d$ and most
excess energy occurs in the form of classical oscillations. This
allows for propagating the wave function on a small \textit{moving
  grid} that extends around the
instantaneous position and momentum expectation values~\cite{SINGER2010}.
The details of our implementation combining the Fourier representation
and a moving grid are described in~\ref{subs:qmprop}.

\subsection{Initial guess voltages} \label{subs:guessgen}
Any optimization, no matter whether it employs classical or quantum
equations of motion, starts from an initial guess. For many
optimization problems, and in particular when using gradient-based methods
for optimization, a physically motivated initial guess is crucial
for success of the optimization~\cite{KochPRA04}. Here,
we design the  initial guess  for the voltage
ramps such that the ion is dragged from
position $x_1$ to $x_2$ in a smooth fashion. This is
achieved as follows:
The trapping potential $V(x,t)$ can be described by the position of
its local minimum $\alpha(t)$. Obviously, $\alpha(t)$ needs
to fulfill the boundary conditions $\alpha(0) = x_1$, $\alpha(T) =
x_2$. In order to ensure smooth acceleration and deceleration of the
center of the trap, we also demand $\dot\alpha(0) =
\dot\alpha(T)=\ddot{\alpha}(0)=\ddot{\alpha}(T)=0$.
A possible ansatz fulfilling these boundary conditions is given by
a polynomial of order 6,
\begin{eqnarray}\label{eqn:transfunc1}
  \alpha(t) = x_1 + d (10 s^3-15s^4+6s^6)\,,
\end{eqnarray}
where $d=x_2-x_1$ denotes the transport distance and $s=t/T$ is a
dimensionless time.

\begin{figure}[tbp]
    \centering
    \includegraphics[width=0.45\textwidth]{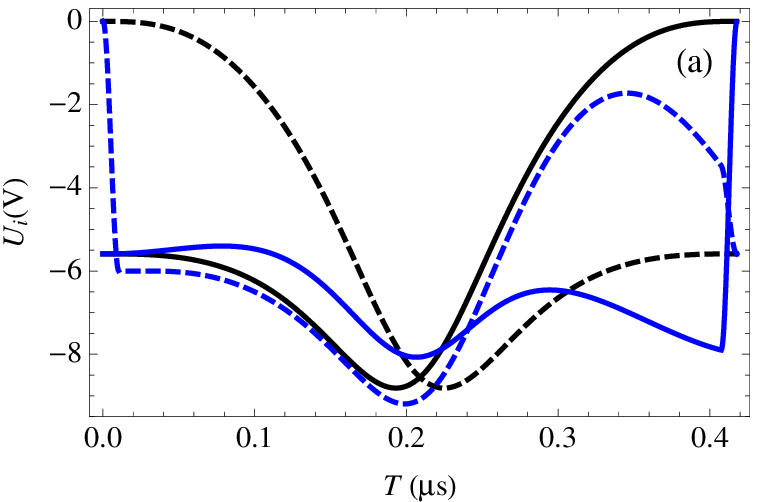}
    \hspace{0.3cm}
    \includegraphics[width=0.45\textwidth]{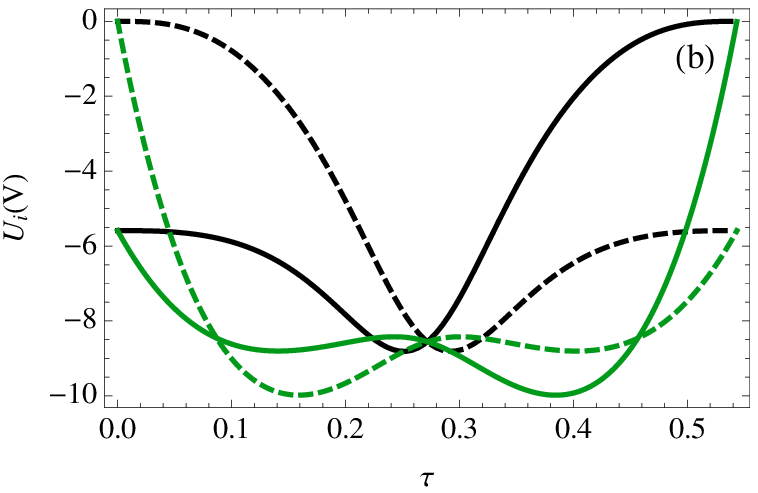}
    \caption{Control voltages applied to electrodes for transporting a $^{40}$Ca$^+$ ion
    from electrode 1 (solid lines) to electrode 2 (dashed lines)
    within 418$\,$ns for an initial trap
    frequency of $\omega=2 \pi \cdot 1.3\,$MHz: Initial guess voltage ramps
    (a and b, black) and ramps obtained by classical optimization (a,
    blue) and the invariant-based inverse engineering (b, green).}
\label{fig:guesscfoct}
\end{figure}

To derive initial guess voltages $U_i^0(t)$,
we use as a first condition that the local minimum of the potential
coincides with $\alpha(t)$. Second, we fix the trap frequency $\omega$
to a constant value throughout the whole shuttling process,
\begin{equation}
\begin{array}{rl}
\left.\frac{\partial V}{\partial x}\right|_{x=\alpha(t)} &=\phi_1'(\alpha(t)) U_1^0(t) + \phi_2'(\alpha(t)) U_2^0(t) \stackrel{!}{=} 0,\\
\left.\frac{\partial^2 V}{\partial x^2}\right|_{x=\alpha(t)} &=\phi_1''(\alpha(t)) U_1^0(t) + \phi_2''(\alpha(t)) U_2^0(t) \stackrel{!}{=} m \omega^2\,.
\end{array}\label{eqn:guesspots}
\end{equation}
These equations depend on first and second order spatial derivatives of the
electrode potentials. Solving for $U_1^0(t)$, $U_2^0(t)$, we obtain
\begin{equation}
U_i^0 (t) = \frac{(-1)^i m \omega ^2  \phi_j'(\alpha(t))}{
\phi_2''(\alpha(t)) \phi_1'(\alpha(t)) - \phi_2'(\alpha(t))\phi_1''(\alpha(t))
}, \quad i,j \in \{1,2\}, \quad j \neq i \label{eqn:solguess}.
\end{equation}
An example is shown in Fig.~\ref{fig:guesscfoct}. If the
electrode potentials have translational symmetry, i.e., $\phi_j(x)=\phi_i(x+d)$,
then $U^0_1(t)=U^0_2(T-t)$. This condition is approximately met for
sufficiently homogeneous trap architectures.
%

\subsection{Geometric optimal control}\label{subsec:geometric}
Most current ion traps are fairly well described by a simple harmonic
model,
\begin{equation}
V(x,t) = -u_1(t) \frac{1}{2}m \omega_0^2 (x-x_1)^2 -
u_2(t) \frac{1}{2}m \omega_0^2 (x-x_2)^2\,,
\end{equation}
where $\omega_0$ is the trap frequency and $u_i$ are dimensionless control
parameters which correspond to the electrode voltages.
Since the equations of motion can be solved analytically, one can also
hope to solve the control problem analytically. One option is given by
Pontryagin's maximum principle~\cite{PONTRY,CHEN2011} which allows to
determine time-optimal controls. Compared to numerical optimization
which always yields local optima, Pontryagin's maximum principle
guarantees the optimum to be global.

In general, the cost functional, 
\begin{equation}
J[\mathbf{u}] = \int_0^T g(\mathbf{y},\mathbf{u}) \dd t\,,
\end{equation}
is minimized
for the equation of motion $\dot{\mathbf{y}} = \mathbf{f}(\mathbf{y},
\mathbf{u})$ and a running cost $g(\mathbf{y},\mathbf{u})$ with
$\mathbf{u} = (u_1, u_2)$ and $\mathbf{y} = (x,v)$ in our
case. The optimization problem is formally equivalent to finding a
classical trajectory by the principle of least action.
The corresponding classical control Hamiltonian that completely
captures the optimization problem is given by
\begin{equation}
H_c(\mathbf{p},\mathbf{y},\mathbf{u}) = p_0 g(\mathbf{y},\mathbf{u}) + \mathbf{p} \cdot \mathbf{f}(\mathbf{y}, \mathbf{u}) \label{eqn:hamcontrol}
\end{equation}
with costate $\mathbf{p}$, obeying
\begin{equation}\label{eq:costate}
\dot{\mathbf{p}}=-\frac{\partial H_c}{\partial \mathbf{y}}\,,
\end{equation}
and $p_0 < 0$ a constant to compensate dimension.
Pontryagin's principle states that $H_c$ becomes maximal for the
optimal choice of $\mathbf u(t)$~\cite{PONTRY,CHEN2011}.

Here we seek to minimize the transport time $T$. The cost functional
then becomes
\[
J[\mathbf{u}] = \int_0^{T_{\rm{min}}} \dd t = T_{\rm{min}}\,,
\]
which is independent of $\mathbf {u}$ itself and leads to
$g(\mathbf{y},\mathbf{u})= 1$. Inserting the classical equations of motion
$\dot{\mathbf{y}} = (v, -\partial_x V)$, the control Hamiltonian becomes
\begin{equation}
  \label{eq:H_c}
  H_c(\mathbf{p},\mathbf{y},\mathbf{u}) = p_0 + p_1 v
  + p_2 \left( u_1 \cdot (x-x_1) + u_2 \cdot (x-x_2)\right) \omega_0^2\,.
\end{equation}
We bound  $u_1$ and $u_2$ by $u_{\rm
max}$ which corresponds to the experimental voltage limit.
Since $H_c$ is linear in $u_i$ and $x_1 \leq x \leq x_2$,
$H_c$ becomes maximal depending on the sign of $p_2$,
\begin{equation}
  u_1(t)= - u_2(t)= \rm{sign}(p_2) u_{\rm max}\,.
\label{eqn:biasmax}
\end{equation}
Evaluating Eq.~\eref{eq:costate} for $H_c$ of Eq.~\eref{eq:H_c} leads to
\begin{eqnarray}
\dot{p_1} = p_2 \omega_0^2 \left(u_1 - u_2\right)\\
\dot{p_2} = -p_1.
\end{eqnarray}
In view of Eq.~\eref{eqn:biasmax}, the only useful choice is $p_2(0) > 0$.
Otherwise the second electrode would be biased to a positive voltage, leading to
a repulsive instead of an attractive potential acting on the ion.
The equations of motion for the costate thus become
\begin{eqnarray}
\dot{p_1} = 0 & \Rightarrow p_1(t) = c_1\\
\dot{p_2} = -p_1& \Rightarrow p_2(t) = p_2(0) - c_1 t.\label{eq:p2}
\end{eqnarray}
For a negative constant $c_1$, $p_2$ is never going to cross
zero. This implies that there will not be a switch in voltages
leading to continuous acceleration. For positive $c_1$ there will be a zero
crossing at time $t_{\rm sw} = p_2(0)/c_1$. The optimal solution thus
corresponds to a single switch of the voltages. We will analyze this
solution and compare it to the solutions obtained by numerical
optimization below in Section~\ref{subs:apcontrol}.

\subsection{Invariant based inverse engineering}\label{subs:invmeth0}
For quantum mechanical equations of motion,
geometric optimal control is limited to very simple dynamics such as
that of three- or four-level systems, see
e.g. Ref.~\cite{HaidongPRA12}.
A second analytical approach that is perfectly adapted to the quantum
harmonic oscillator
utilizes the Lewis-Riesenfeld theory which introduces dynamical
invariants and their eigenstates~\cite{lewis1969}. This
invariant-based inverse engineering approach (IEA) has recently been
applied to the transport problem~\cite{TORR2011,PalmeroPRA13}. The basic idea is
to compensate the inertial force occurring during the transport
sequence. To this end, the potential is written in the following form:
\begin{equation}\label{eqn:invpot}
V(x,t) = -F(t) x  + \frac{m}{2} \Omega^2(t) x^2 + \frac{1}{\rho^2(t)} U\left(\frac{x-\alpha(t)}{\rho(t)}\right)\,.
\end{equation}
The functions $F$, $\Omega$, $\rho$ and $\alpha$ have to fulfill
constraints,
\begin{eqnarray}
\ddot \rho(t) + \Omega^2(t)\rho(t) = \frac{\Omega_0^2}{\rho^3(t)}\,,\\
\ddot \alpha(t) + \Omega^2(t)\alpha(t) = F(t)/m\,, \label{eqn:compforce0}
\end{eqnarray}
where $\Omega_0$ is a constant and $U$ an arbitrary function. We choose
$\Omega(t)=\Omega_0=0$, $\rho(t) = 1$, and $\alpha(t)$ to be the
transport function of Sec.~\ref{subs:guessgen}. This enables us to
deduce the construction rule for $F(t)$, using Eq.~\eref{eqn:compforce0},
\begin{equation}\label{eq:F}
\ddot\alpha(t) = F(t)/m\,,
\end{equation}
such that $F(t)$ compensates the inertial force given by the
acceleration of the trap center.
For the potential of Eq.~\eref{eqn:invpot}, the Hermitian operator
\begin{equation}
\Op{I} = \frac{1}{2m}
\left[\rho\left(p-m\dot\alpha\right)-m\dot\rho\left(x-\alpha\right)\right]^2
+\frac{1}{2} m \Omega_0^2 \left(\frac{x-\alpha}{\rho}\right)^2
+ U\left(\frac{x-\alpha}{\rho}\right)
\end{equation}
fulfills the invariance condition for all conceivable quantum states $\ket{\Psi(t)}$:
\begin{equation}
\frac{\rmd }{\rmd t }\braket{\Psi(t)|\Op{I}(t)|\Psi(t)} = 0 \quad \Leftrightarrow \quad \frac{\rmd \Op{I}}{\rmd t } = \frac{\partial \Op{I} }{\partial t} + \frac{1}{\rmi \hbar} [\Op{I}(t),\Op{H}(t)] = 0
\end{equation}
with $\Op{H}$ the Hamiltonian of the ion.
The requirement for transporting the initial ground state to the
ground state of the trap at the final time corresponds to $\Op{H}$ and
$\Op{I}$ having a common set
of eigenfunctions at initial and final time. This is the case for
$\dot\alpha(0)= \dot\alpha(T) = \dot{\rho}(t) =0$~\cite{DHARA1984,TORR2011}.
We can now identify $U$ in Eq.~\eref{eqn:invpot} with the trapping
potential of Eq.~\eref{eq:V}.
The additional compensating force
is generated using the same trap electrodes
by applying an additional voltage $\delta U_i$. For a given
transport function $\alpha(t)$ we therefore have to solve the underdetermined
equation,
\begin{equation}
m \ddot{\alpha}(t) = -\phi_1'(x(t)) \delta U_1(t) - \phi_2'(x(t)) \delta U_2(t),\label{eqn:regul}
\end{equation}
where $x(t)$ is given by the classical trajectory. Since the ion
is forced to follow the center of the trap we can set
$x(t)=\alpha(t)$. The compensating force is supposed to be a function
of time only, cf. Eq.~\eref{eq:F}, whereas 
changing the electrode voltages by $\delta U_i$ will, via the
$\phi_i(x)$, in general yield a 
position-dependent force. This leads to a modified second derivative
of the actual potential:
\begin{equation}
m \omega_c(t)^2 =\sum_{i=1}^2\phi_i''(\alpha(t))(U_i^0(t)+\delta U_i(t)) = m( \omega^2 + \delta\omega(t)^2)\,,
\end{equation}
where $\delta\omega(t)^2$ denotes the change in trap frequency due to the
compensation voltages $\delta U_i$, $\omega$ is the initially desired trap
frequency, and $U_i^0(t)$ is found in Eq.~\eref{eqn:solguess}. 
A time-varying actual frequency $omega_c(t)$ might lead to wavepacket
squeezing.  
However, since Eq.~\eref{eqn:regul} is underdetermined, we can set
$\delta\omega(t)^2 = 0$ leading to $\omega_c(t) = \omega$ as desired. 
With this condition we can solve Eq.~\eref{eqn:regul} and obtain
\begin{equation}\label{eqn:compconstw}
\delta U_i (t) = \frac{\ddot \alpha(t) \,(-1)^i \, m \,\phi_j''(\alpha(t))}{
\phi_2''(\alpha(t)) \phi_1'(\alpha(t)) - \phi_2'(\alpha(t))\phi_1''(\alpha(t))
}\,,~i,j \in \{1,2\}\,,~j \neq i\,.
\end{equation}
Note that Eq.~\eref{eqn:compconstw} depends only on the trap
geometry. The transport duration
$T$ enters merely as a scaling parameter via $\ddot \alpha(t) =
\alpha''(s)/T^2$. An example
of a voltage sequence obtained by this method is shown in
\fref{fig:guesscfoct}(b). The voltage curves are symmetric under time
inversion like the guess voltages, that are derived from the same potential
functions $\phi_i(x)$.

\section{Application and comparison of the control methods}\label{subs:apcontrol}
We now apply the control strategies introduced in Sec.~\ref{subs:ttcp}
to a scenario with the parameters chosen to correspond to a
typical experimental setting. The
scaling of the classical speed limit is studied for a fixed maximum control
voltage range and we show how in the limiting case the \textit{bang-bang} solution is
obtained. To verify the validity of the classical solution we are
applying the obtained voltage ramps to a quantum mechanical wave
packet propagation. Similarly, we use the invariant-based approach and
verify the result for a quantum mechanical propagation.

\subsection{Experimental constraints and limits
  to control  for classical ion transport}\label{subs:kloct2}

\begin{figure}[tbp]
\centering
\includegraphics[width=0.45\textwidth]{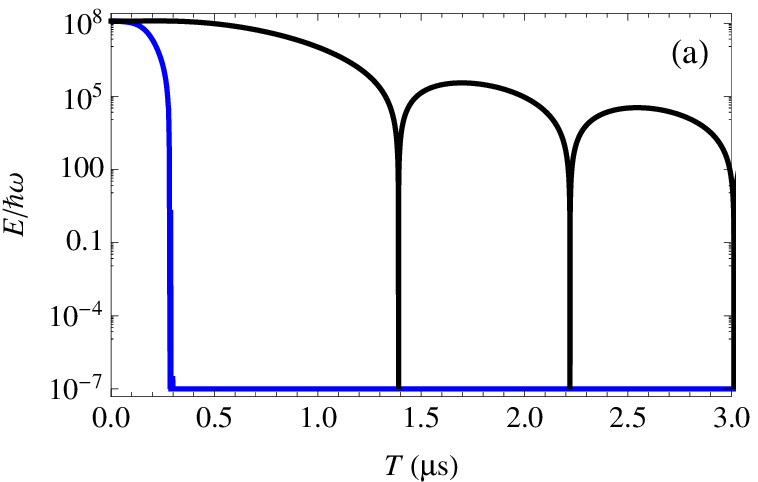} 
\includegraphics[width=0.45\textwidth]{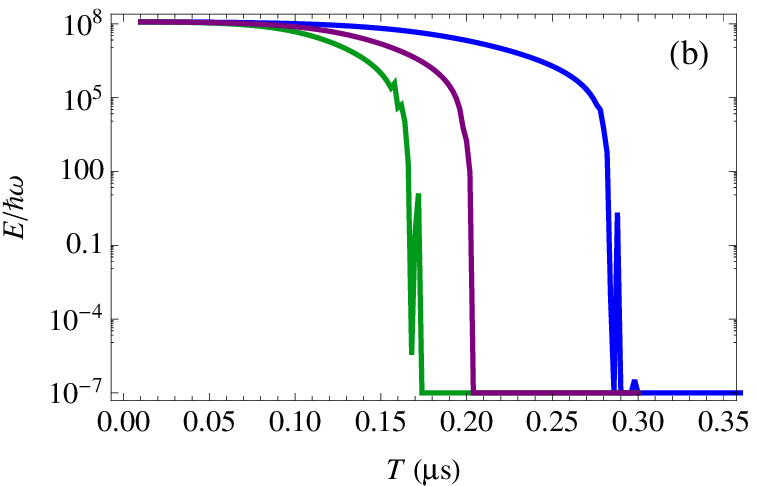} 
\caption{Final energy vs. transport time for different voltage
  ramps and classical dynamics. (a) shows the improvement over the
  initial guess (black) by numerical optimization for a
  maximum voltage of 10$\,$V (blue) and (b) compares the results of
  numerical optimization for maximum voltages of 10$\,$V (blue),
  20$\,$V (purple), and 30$\,$V (green). The spikes in (b) are due to
  voltage truncation.
}
\label{fig:transportopts}
\end{figure}
\begin{figure}[tbp]
  \begin{center}
    \includegraphics[width=0.44\textwidth]{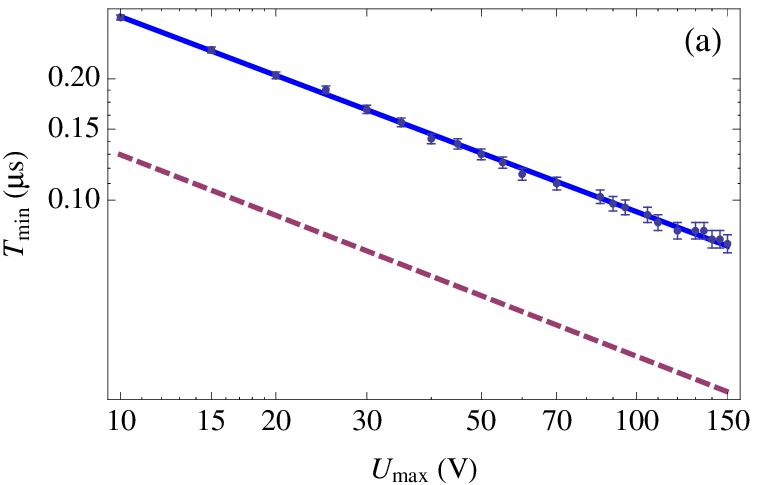}
    \hspace{0.3cm}
    \includegraphics[width=0.45\textwidth]{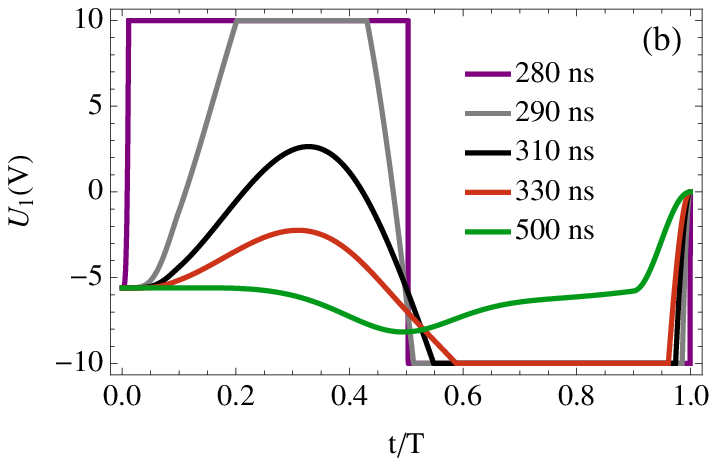}
    \caption{(a) Minimum transport time $T^\mathrm{opt}_{\rm min}$ vs.
      maximum electrode voltage $U_{\rm max}$, obtained from numerical
      optimization of classical transport dynamics (blue dots)
      along with a
      fit to Eq.~\eref{eqn:fitfn}. We also indicate the analytic
      bang-bang result, Eq.~\eref{eqn:Ttheo}, derived for idealized,
      purely harmonic potentials (purple dashed line) being proportional to
      $1/\sqrt{U_{\rm{max}}}$.
      The slopes of the curves are clearly similar,
      indicating the negligible impact of anharmonicities on the
      scaling of $T^\mathrm{opt}_{\rm min}$ with $U_{\rm max}$.
      (b) Optimized voltages for the left electrode
      with $U_{\rm{max}}=10\,$V: The
      shorter the transport time, the more the ramp approaches a
      square shape. The rectangular
      bang-bang-like solution is attained at $T=280\,$ns, where the
      classical control of energy neutral transport breaks down due to
      an insufficient voltage range.}
\label{fig:classlimit}
\end{center}
\end{figure}
In any experiment, there is an upper limit to the electrode voltages
that can be applied. It is the range of electrode voltages that limits
the maximum transport speed.
Typically this range is given by $\pm$10$\,$V for technical reasons. It
could be increased by the development of better voltage
supplies. We define the minimum possible transport time $T_{\rm
  {min}}$ to be the smallest time $T$ for which
less than $0.01$ phonons are excited due to the total transport.
To examine how $T_{\rm min}$ scales as a
function of the maximum electrode voltages $U_{\rm max}$, we
have carried out numerical optimization combined with classical
equations of motion. The initial guess voltages, cf.
Eqs.~\eref{eqn:transfunc1} and~\eref{eqn:solguess}, were taken to
preserve a constant trap frequency of $\omega = 2 \pi \cdot 1.3$~MHz for
a $^{40}\rm{Ca}^+$ ion. The transport ramps were optimized for a
range of maximum voltages between 10-150~V and transport times between
10 ns and 300 ns with voltages truncated to
$\pm\,U_{\rm max}$ during the updates. The results are shown in
Figs.~\ref{fig:transportopts} and~\ref{fig:classlimit}.
Figure~\ref{fig:transportopts} depicts the final excitation energy
versus transport time, comparing the initial guess (black) to an
optimized ramp with $U_{\rm max}=10\,$V (blue) in
Fig.~\ref{fig:transportopts}(a). For the initial guess, the final
energy displays an oscillatory behavior with respect to the
trap period ($T_{\rm{per}}=0.769\,\mu$s for $\omega = 2 \pi \cdot 1.3\,$MHz)
as it has been experimentally observed in
Ref.~\cite{WALTHER2012}, and an overall decrease of the final energy
for longer transport times. The optimized transport with
$U_{\rm max}=10\,$V (blue line in Fig.~\ref{fig:transportopts}(a))
shows a clear speed up of energy neutral transport:
An excitation energy of less than 0.01 phonons is obtained for
$T^\mathrm{opt}_\mathrm{min}=0.284\,\mu$s  compared to $T^\mathrm{guess}_\mathrm{min}=1.391\,\mu$s.
The speedup increases with maximum voltage as shown in
Fig.~\ref{fig:transportopts}(b).
The variation of $T^\mathrm{opt}_{\rm min}$ on $U_{\rm max}$ is
studied in Fig.~\ref{fig:classlimit}(a). We find
a functional dependence of
\begin{equation}
  T^\mathrm{opt}_{\rm min}(U_{\rm max}) \approx
  a \left(\frac{U_{\rm max}}{1\,{\rm V}}\right)^{-b}
\label{eqn:fitfn}
\end{equation}
with $a = 0.880(15)\,\mu$s and $b = 0.487(5)$.
Optimized voltages are shown in Fig.~\ref{fig:classlimit}(b)
for the left electrode with $U_{\rm{max}}=10\,$V. As the transport
time decreases, the voltage ramp approaches a square shape. A
bang-bang-like solution is attained at $T=280\,$ns. However, for such
a short transport time,
classical control of energy neutral transport breaks down due to
an insufficient voltage range and the final excitation amounts to 5703
mean phonons.

In the following we show that for purely harmonic potentials, the
exponent $b$ in Eq.~\eref{eqn:fitfn} is universal, i.e., it does not depend
on trap frequency nor ion mass. It is solely determined by the
bang-bang like optimized voltage sequences, where instantaneous
switching between maximum acceleration and deceleration guarantees
shuttling within minimum time. The technical feasibility of bang-bang
shuttling is thoroughly analyzed in Ref.~\cite{ALONSO2013}.
The solution is obtained by the application of
Pontryagin's maximum principle \cite{PONTRY,CHEN2011} as discussed in
Sec.~\ref{subsec:geometric} and assumes instantaneous
switches. Employing Eqs.~\eref{eqn:biasmax} and~\eref{eq:p2},
the equation of motion becomes
\begin{equation}
\ddot x = \omega_0^2 u_{\rm max} \cdot \left\{
\begin{array}{cc}
d , & t < t_{\rm sw}\\ -d, & t > t_{\rm sw}
\end{array} \right. .
\end{equation}
This can be integrated to
\begin{equation}
x(t)=  \left\{
\begin{array}{lc}
x_1 + u_{\rm max} d \omega_0^2t^2 , & 0 \leq t \leq t_{\rm sw}\\
x_1 + d - u_{\rm max} d \omega_0^2(t-T_{\rm{min}})^2 , & t_{\rm sw} \leq t \leq T_{\rm{min}}
\end{array} \right.
\end{equation}
with the boundary conditions $x(0) = x_1$, $x(T_{\rm{min}}) = x_2$ and $\dot{x}(0) =
\dot{x}(T_{\rm{min}}) = 0$. Using the continuity of $\dot x$ and $x$
at $t=t_{\rm sw}$, we obtain
\begin{equation}
t_{\rm sw} = \frac{T}{2} ,\quad T_{\rm{min}} = \frac{\sqrt{2}}{\omega_0} \sqrt{\frac{1}{u_{\rm max}}}. \label{eqn:Ttheo}
\end{equation}
Notably, the minimum transport time is proportional to $u_{\rm
max}^{-1/2}$ which explains the behavior of the numerical data 
shown in Fig.~\ref{fig:classlimit}.
This scaling law can be understood intuitively by considering that in the bang-bang
control approach, the minimum shuttling time is given by the shortest attainable
trap period, which scales as $u_{max}^{-1/2}$.
Assuming a trap frequency of $\omega_0=2 \pi \cdot 0.55$~MHz  in
Eq.~\eref{eqn:Ttheo}, corresponding to a trapping voltage of $-1\,$V
for our trap geometry, we find a prefactor $\sqrt{2}/\omega_0 =
0.41 \mu$s. This is smaller than $a=0.880(15)\mu$s obtained by
numerical optimization for realistic trap potentials. The difference
can be rationalized in terms of the average acceleration
provided by the potentials. For realistic trap geometries, the force
exerted by the electrodes is inhomogeneous along the transport path.
Mutual shielding of the electrodes reduces the electric field
feedthrough of an electrode to the neighboring ones. Thus, the
magnitude of the accelerating force that a real electrode can
exert on the ion 
when it is located at a neighboring electrode is reduced with respect
to a constant force generating harmonic potential with the same trap
frequency.

The minimum transport time of $T^\mathrm{opt}_\mathrm{min}=0.284\,\mu$s 
identified here for $U_\mathrm{max}=10\,$V, cf. the blue line in
Fig.~\ref{fig:transportopts}(a), is significantly shorter than 
operation times realized experimentally. For comparison, an ion has
recently been shuttled within $3.6\,\mu$s, leading to a final
excitation of $0.10\pm0.01$ motional quanta~\cite{WALTHER2012}. 
Optimization may
not only improve the transport time but also the stability with
respect to uncertainties in the time. This is in contrast to the 
extremely narrow  minima of the final excitation energy for the guess
voltage ramps shown in black in Fig.~\ref{fig:transportopts}(a),
implying a very high
sensititivity to uncertainties in the transport time. For example, for
the fourth minimum of the black curve, located at $3.795\,\mu$s and close
to the operation time of Ref.~\cite{WALTHER2012} (not shown in
Fig.~\ref{fig:transportopts}(a)), final excitation energies of less
than  0.1 phonons are observed only within a window of 3$\,$ns.
Optimization of the voltage ramps for $T=3.351\,\mu$s increases
the stability against variations in transport time to 
more than $60\,$ns.

In conclusion we find that optimizing the classical motion of an ion 
allows us to identify the
minimum operation time for a given maximum voltage and improve the
stability with respect to timing uncertainies for longer operation
times. The analytical solution derived from Pontryagin's maximum
principle is helpful to understand the minimum time control strategy.
Numerical optimization accounts for all typical features of realistic
voltage ramps. It allows 
for identifying the minimum transport time, predicting $36.9\%$ of the
oscillation period for current maximum voltages and a trap frequency of
$\omega = 2 \pi\cdot 1.3\,$MHz. This number can be
reduced to $12.2\%$  when increasing the maximum voltage by one order of
magnitude.

However, these predictions may be rendered invalid by a breakdown of
the classical approximation.

\subsection{Validity of classical solutions in the
quantum regime}\label{subs:qmprop2}
\begin{figure}[tbp]
\centering
\includegraphics[width=0.45\textwidth]{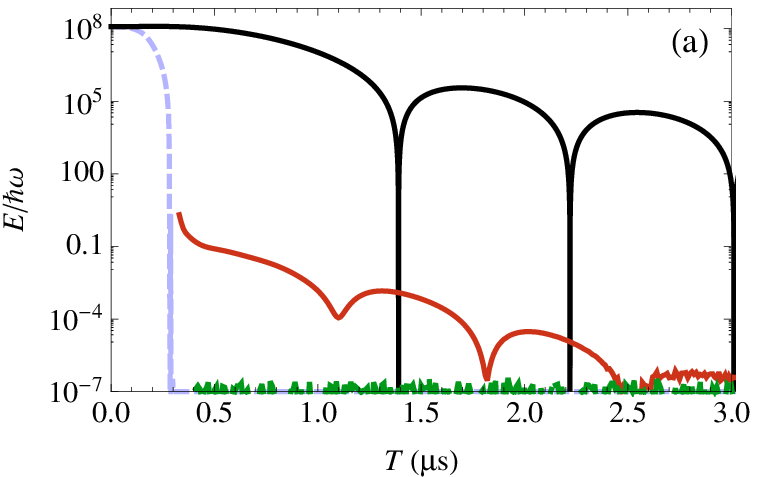}
\includegraphics[width=0.43\textwidth]{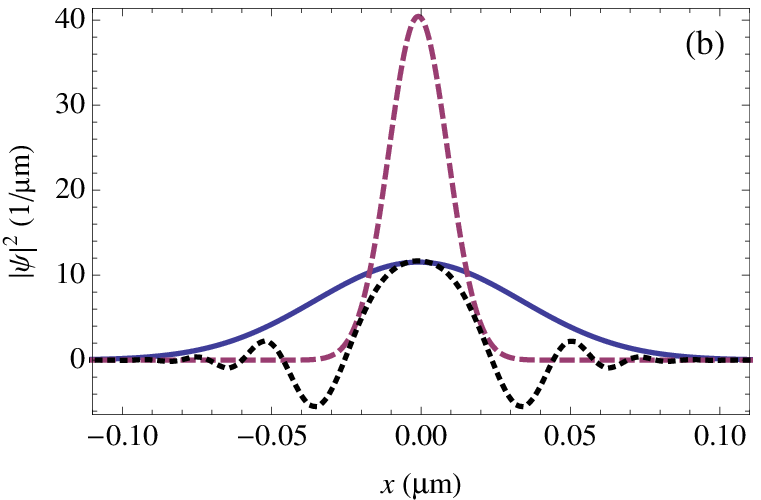}
\caption{Testing control strategies obtained with classical dynamics
  for wavepacket motion:
  (a) Final excitation energy of the ion wavepacket
  with the initial guess (black) and the optimized voltage ramps with
  $U_\mathrm{max}=10\,$V (red). Also shown is a solution obtained by
  invariant based inverse engineering for a quantum mechanical
  harmonic oscillator (green). For comparison, the final excitation
  energy obtained by solving the classical equation of motion
  with the optimized ramp is shown in light blue. Note that for the
  initial guess (black), the relative difference between wavepacket
  and classical motion is not visible on the scale of the figure (less
  than $10^{-3}$).
  (b) Final wavefunction $|\Psi(T)|^2$ (blue) for classically optimized
  transport with $T = 320\,$ns and $U_\mathrm{max}=10\,$V and target wave
  function $|\Psi^{\rm tgt}|^2$ (purple dashed). Also shown is the scaled real
  part of the final wavefunction $\Re(\Psi(T))$ (black dotted). The
  clearly visible spread of the wavepacket corresponds to squeezing of
  the momentum.}
 \label{fig:sqlimit}
\end{figure}
We now employ quantum wavepacket dynamics to test the classical
solutions, obtained in Sec.~\ref{subs:kloct2}.
Provided the trap frequency is constant  and the trap is perfectly
harmonic, the wave function will only be displaced during the
transport. For a time-varying trap frequency, however, squeezing may
occur~\cite{scu11}. In extreme cases, anharmonicities of the potential
might lead to wavepacket dispersion. Since these two
effects are not accounted for by numerical optimization of classical
dynamics, we discuss in the following at which timescales such
genuine quantum effects become significant. To this end, we have
employed the optimized voltages shown in Fig.~\ref{fig:classlimit}(b)
in the propagation of a quantum wavepacket. We compare the results of
classical and quantum mechanical motion in Fig.~\ref{fig:sqlimit}(a),
cf. the red and lightblue lines. A clear deviation is observed.
Also, as can be seen Fig.~\ref{fig:sqlimit}(b), the
wavefunction fails to reach the target wavefunction for transport
times close to the classical limit $T^\mathrm{opt}_{\rm{min}}$. This is
exclusively caused  by squeezing and can be verified by inspecting
the time evolution of the wavepacket in the final potential: We find the
width of the wavepacket to oscillate, indicating  a squeezed
state. No wavepacket dispersion effects are observed, i.e., the final
wavepackets are still minimum uncertainty states, with $\min(\Delta
x\cdot\Delta p) = \hbar/2$. This means that no effect of
anharmonicities in the potential is observed.
An impact of anharmonicities is expected once the size of the
wavefunction becomes comparable to  segment distance $d$ (see
Fig.~\ref{fig:electrodes}). Then the wavefunction extends over
spatial regions in which the potentials
deviate substantially from harmonic potentials.
For the ion shuttling problem, this effect does not play a role over the
relevant parameter regime.
The effects of anharmonicities in
the quantum regime for trapped ions were thoroughly analyzed in
Ref.~\cite{HOME2011}.
Squeezing increases $T_{\rm min}$ from $0.28\,\mu$s to $0.86\,\mu$s
for the limit of exciting less than 0.01 phonons, see the red curve in 
Fig.~\ref{fig:sqlimit}(a), i.e., it only triples the minimum
transport time.
We show that squeezing can be suppressed altogether in the following
section. 

\subsection{Application of a compensating force approach}\label{subs:invmeth}
\begin{figure}[tbp]
\begin{center}
\includegraphics[width=0.5\textwidth]{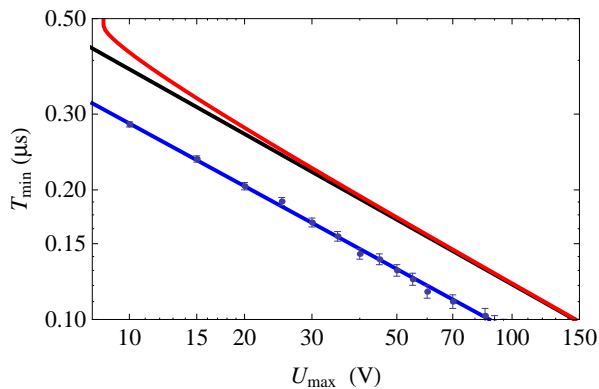}
\caption{Minimum transport time $T_{\rm min}$ vs. maximum electrode
  voltage $U_{\rm max}$ obtained by the invariant-based inverse
  engineering approach: The compensating force method for a trap
  frequency of $\omega = 2 \pi\cdot 1.3\,$MHz using
  the transport function of Eq.~\eref{eqn:transfunc1}
  (red) and for
  the limiting case of vanishing trap frequency (black). Also
  shown is the classical result from Fig.~\ref{fig:classlimit} (blue).}
\label{fig:compforcelimit}
\end{center}
\end{figure}
In the invariant-based IEA, the minimal transport time is determined by the
maximum voltages that are required for attaining zero motional
excitation. The total voltage that needs to be applied is given by
$U_i(t)=U_i^0(t)+\delta U_i(t)$ with $U_i^0(t)$ and $\delta U_i(t)$
found in Eqs.~\eref{eqn:solguess} and~\eref{eqn:compconstw}.
The maximum of $U_i(t)$, and thus the mininum in $T$, 
is strictly 
related to the acceleration of the ion provided by the transport
function $\alpha(t)$, cf. Eq.~\eref{eqn:compconstw}.
If the acceleration is too high, the voltages
will exceed the feasibility limit $U_{\rm max}$.
At this point it can also be understood why the acceleration should be zero
at the beginning and end of the transport: For $\ddot\alpha(0)\neq 0$
a non-vanishing correction voltage
$\delta U_i \neq 0$ is obtained from Eq.~\eref{eqn:compconstw}. This
implies that the voltages do not match the initial trap conditions,
where the ion should be located at the center of the initial
potential.

We can derive a transport function $\alpha(t)$ compliant with the
boundary conditions using Eq.~\eref{eqn:transfunc1}.
For this case,
Fig.~\ref{fig:compforcelimit} shows the transport time
$T^\mathrm{IEA}_{\rm{min}}$ versus the maximum voltage $U_{\rm{max}}$
that is applied to the electrodes during the transport sequence.
For large transport times, the initial guess voltages $U_i^0(t)\propto\omega^2$
dominate the compensation voltages $\delta U_i(t)\propto\ddot \alpha(t) =
\alpha''(s)/T^2 $. This leads to the bend of the red curve. When the trap
frequency $\omega$ is lowered, the bend decreases. For the 
limiting case of no confining potential $\omega=U_i^0(t)=0$,
$T^\mathrm{IEA}_\mathrm{min}$ is solely determined by the compensation
voltages.   
In this case the same scaling of $T^\mathrm{IEA}_\mathrm{min}$ with
$U_\mathrm{max}$ as for
the optimization of classical dynamics is observed, cf. black and blue
lines in Fig.~\ref{fig:compforcelimit}. For large $U_\mathrm{max}$,
this scaling also applies to the case of non-zero trap frequency,
cf. red line in Fig.~\ref{fig:compforcelimit}.

We have tested the performance of the compensating force by employing
it in the time evolution of the wavefunction. It leads to near-perfect
overlap with the target state with an infidelity of less than
$10^{-9}$. The final excitation energy of the propagated wave function
is shown in Fig.~$\ref{fig:sqlimit}$ (green line) for a maximum voltage of
$U_{\rm max} =10\,$V. For the corresponding minimum transport time,
$T^\mathrm{IEA}_{\rm  min}(10\,\rm{V}) = 418\,$ns,  a final
excitation energy six orders of magnitude below that found by
optimization of the classical dynamics is obtained. This demonstrates
that the invariant-based IEA is capable of avoiding the wavepacket
squeezing that was observed in Sec.~\ref{subs:qmprop2}
when employing classically optimized 
controls in quantum dynamics. 
It also confirms that anharmonicities do
not play a role since these would not be accounted for by the
IEA-variant employed here. Note that an adaptation of the
invariant-based IEA to anharmonic traps is found in
Ref.~\cite{PalmeroPRA13}. 
Similarly to numerical optimization of classical dynamics, 
IEA is capable of improving the
stability against variations in transport time $T$.
The final excitation energy obtained for $T=3.351\,\mu$s stays below
0.1 phonons within a window of more than $13\,$ns.

A further reduction of the minimum transport time may be achieved
due to the freedom of choice in the transport function
$\alpha(t)$, by employing higher polynomial orders in order
to reduce the compensation voltages $\delta U_i(t)$, cf.
Eq.~\eref{eqn:compconstw}. 
However, the fastest quantum mechanically valid transport
has to be slower than the solutions obtained for 
classical ion motion. This follows from the bang-bang control being
the time-optimal solution for a given voltage limit and the IEA
solutions requiring additional voltage to compensate the wavepacket
squeezing. 
We can thus conclude that the time-optimal quantum solution
will be inbetween the blue and black curves of
Fig.~\ref{fig:compforcelimit}. 

\subsection{Feasibility analysis of quantum optimal control}\label{subs:qmoct2}
Numerical optimization of the wavepacket motion is expected to become
necessary once the dynamics explores spatial regions in which the
potential is strongly anharmonic or is subject to strongly anharmonic
fluctuations. This can be expected, for example, when the spatial
extent of the wavefunction is not too different from that of the
trap. Correspondingly, we introduce the parameter $\xi=\sigma_0/d$,
which is the wavefunction size normalized to the transport distance.
While for current trap
architectures, such a scenario is rather unlikely, further
miniaturization might lead to this regime. Also, it is currently
encountered in the transport of neutral atoms in tailored optical
dipole potentials~\cite{Ivanov2010,WaltherTwo2012}.

Gradient-based quantum OCT requires an initial guess voltage that
ensures a finite overlap of the propagated wave function $\Psi(T)$
with the target state $\Psi^{\rm tgt}$, see
Eq.~\eref{eqn:chioft}. Otherwise, the amplitude
of the co-state $\chi$ vanishes. The overlap can also be
analyzed in terms of phase space volume.
For a typical ion trap setting with parameters as in
Fig.~\ref{fig:electrodes}, the total covered phase space volume in
units of Planck's constant is $ m\,d^2\, \omega /2 \pi h \approx
10^7$. This leads to very slow convergence of the optimization
algorithm, unless an extremely good initial
guess is available.

\begin{figure}
\begin{center}
\includegraphics[width=0.45\textwidth]{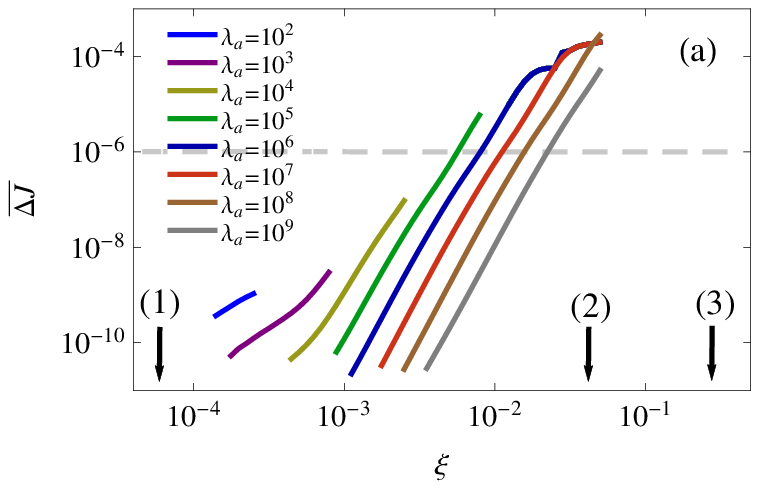}
\includegraphics[width=0.45\textwidth]{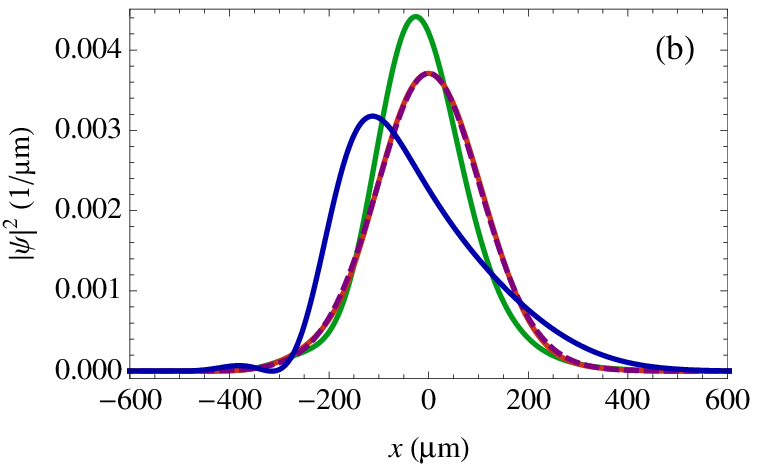}
\caption{(a) Mean improvement of the optimization functional,
  $\overline{\Delta J}$ (averaged over 100   iterations),
  versus relative size of the wavepacket $\xi$ for different
  optimization weights $\lambda_a$, cf. Eq.~\eref{eqn:krotovupdate},
  ranging from $\lambda_a=10^{2}$ (leftmost line) to
  $\lambda_a=10^{9}$ (rightmost line) in powers of ten.
  The arrows indicate: (1) the parameters corresponding to current
  trap technology, (2) good convergence of quantum OCT with the
  invariant-based IEA still being valid, (3) fast convergence of
  quantum OCT with invariant-based IEA starting to fail.
  (b) Final wavefunction amplitudes  for $\xi = 0.4$ (arrow (3) in
  (a))  and classical optimization (blue, fidelity of $83.8\%$), IEA
  (green, $94.6\%$), quantum OCT (red, fidelity of $99.9\%$). Also
  plotted is the target state (purple dashed).}
\label{fig:qmconv}
\end{center}
\end{figure}
We utilize the results of the optimization for
classical dynamics of Sec.~\ref{subs:kloct2} as initial guess
ramps for optimizing the wavepacket dynamics and investigate the
convergence rate as a function of the system
dimension, i.e., of $\xi$. The results are shown in
Fig.~\ref{fig:qmconv}(a), plotting the mean improvement per
optimization step, $\Delta J$, averaged over 100 iterations, versus the
scale parameter $\xi$.
We computed the convergence rate $\overline{\Delta J}$ for different,
fixed optimization weights $\lambda_a$ in
Eq.~\eref{eqn:krotovupdate}.  The curves in Fig.~\ref{fig:qmconv}(a)
are truncated for large values of $\overline{\Delta J}$, where the
algorithm becomes numerically unstable.  Values below
$\overline{\Delta J}=10^{-6}$ (dashed grey line in
Fig.~\ref{fig:qmconv}(a)) indicate an insufficient convergence rate
for which no significant gain of fidelity is obtained with
reasonable computational resources. In this case the potentials are
insufficiently anharmonic to provide \textit{quantum} control of the
wavefunction.

\begin{figure}[tb]
  \begin{center}
  \begin{minipage}{0.30\textwidth}
  \centering
  \includegraphics[width=\textwidth]{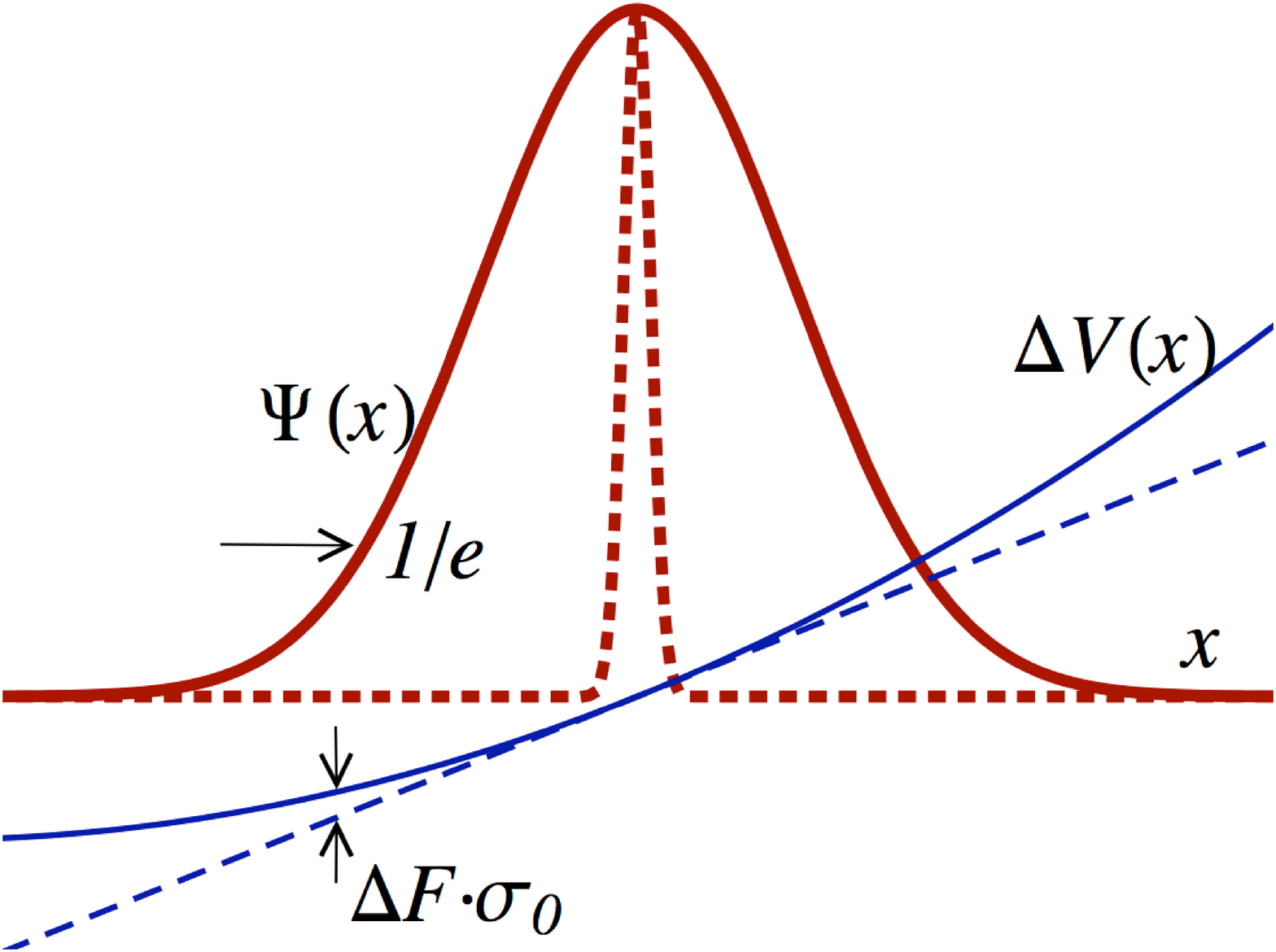}
  \end{minipage}\hspace*{3ex}
  \vspace{0cm}
  \begin{minipage}{0.55\textwidth}
    \begin{center}
    \begin{tabular}
      { l c l l l l l}
      \br
      &$\xi$&$\Delta F/F$ &
      $\mathfrak{F}_{\rm{IEA}}$&
      $\mathfrak{F}_{\rm{qOCT}}$&
      $T_{\rm {CPU}} (h)$\\
      \mr
      (1)&$5.0\cdot10^{-5}$&$2.7\cdot10^{-13}$ & 1.000 &
       N/A & N/A\\
      (2)&$5.0\cdot10^{-2}$&$2.7\cdot10^{-4}$ & 1.000 &
       0.999 & 32.2 \\
      (3)&$4.0\cdot10^{-1}$&$1.2\cdot10^{-1}$ & 0.946 &
       0.999 & 4.3\\
      \br
    \end{tabular}
  \end{center}
  \end{minipage}
  \caption{Limitation of the compensating force approach. A force
    inhomogeneity $\Delta F =
    \sum_i[\phi_i'(\alpha(t)+\sigma_0)-\phi'(\alpha(t)-\sigma_0)]\delta
    U_i(t)$ across the wavefunctions is caused by anharmonicities of
    the potential $\Delta V = F(t)\, x$ used to implement the
    compensating force. The relative spread of the force
    $\Delta F/F$ across the wavefunction is taken at the point in
    time, where the acceleration $\ddot \alpha(t)$ is maximal. $\Delta
    F/F$ increases to the range of several percent for large
    wavefunction extents. This leads to a drop in the fidelity
    $\mathfrak{F}_{\rm{IEA}}$. Also shown is the fidelity
    $\mathfrak{F}_{\rm{qOCT}}$ obtained by optimizing the quantum dynamics.
    The CPU time $T_{\rm{CPU}}$ required for optimization could be
    easily reduced by a factor of 8 in case (3) compared to case (2) due to the
    better convergence of quantum OCT in this regime.}
\label{fig:schema}
  \end{center}
\end{figure}

Numerical optimization of the wavepacket dynamics
is applicable and useful for scale parameters of $\xi\approx
0.05$ and larger, indicated by arrows (2) and (3) in
Fig.~\ref{fig:qmconv}(a). Then the  wavefunction size becomes
comparable to the transport distance, leading for example to a phase
space volume of around $10\,h$ for arrow (2). At this scale the force
becomes inhomogeneous across the wavepacket. This  leads to  a
breakdown of the IEA, as
illustrated for  $\xi  = 0.4$ in Figs.~\ref{fig:qmconv}(b)
and~\ref{fig:schema}.
The fidelity  $\mathfrak{F}_{\rm{IEA}}$ for the IEA drops below
$94.6\%$, whereas $\mathfrak{F}_{\rm{qOCT}}=0.999$ is achieved by
numerical optimization of the quantum dynamics.

\section{Summary and Conclusions}\label{sec:concl}
Manipulation of motional degrees of freedom is very widespread in
trapped-ion experiments. However, most theoretical calculations
involving ion transport over significant distances are based on
approximations that in general do not guarantee the level of precision
needed for high-fidelity quantum control, especially in view of
applications in the context of quantum technologies. As a consequence,
before our work little was known about how to apply optimal control
theory to large-scale manipulation of ion motion in traps, concerning
in particular the most efficient simulation and control methods to be
employed in different parameter regimes, as well as the level of
improvement that optimization could bring.

With this in mind, in the present work we have investigated the
applicability of several classical and 
quantum control techniques for the problem of moving an ion across a
trap in a fast and accurate way. When describing the ion dynamics
purely classically, numerical optimization yields transport times
significantly shorter than a trapping period. The minimum transport
duration depends on the maximal electrode voltage that can be applied
and was found to scale as $1/\sqrt{U_{\rm{max}}}$. The same scaling is
observed for  time-optimal bang-bang-like solutions that can be
derived using Pontryagin's maximum principle and assuming perfectly
harmonic traps. Not surprisingly, the classically optimized solutions
were found to fail when tested in quantum wavepacket motion for
transport durations of about one third of a trapping period. Wavepacket
squeezing turns out to be the dominant source of error with the
final wavepacket remaining a minimum uncertainty state.
Anharmonic effects were found to play no significant role for single-ion
shuttling over a wide range of parameters.
Wavepacket squeezing can be perfectly compensated by the control strategy
obtained with the invariant-based inverse engineering approach. It
amounts to applying correction voltages which can be generated by the
trapping electrodes and which exert a compensating force on the
ion. This is found to be the method of choice for current experimental
settings.

Control methods do not only allow to assess the minimum time required
for ion transport but can also yield more robust solutions. 
For transport times that have been used in recent
experiments~\cite{WALTHER2012}, significantly larger than the minimum
times identified here, the classical solutions are valid also for the 
quantum dynamics. In this regime, both numerical optimization of
classical ion motion and the inverse engineering approach yield a
significant improvement of stability against uncertainties in
transport time. Compared to the initial guess voltages, the time
window within which less than 0.1 phonons are excited after transport
is increased by a factor of twenty for numerical optimization and 
a factor of five for the inverse engineering approach. 

Further miniaturization is expected to yield trapping potentials where
the wavepacket samples regions of space in which the potential, or
potential fluctuations, are strongly anharmonic. Also, for large
motional excitations recent experiments have shown 
nonlinear Duffing oscillator behavior \cite{AKERMAN2010}, nonlinear
coupling of modes in linear ion crystals \cite{ROOS2008,nie2009theory}
and amplitude dependent modifications of normal modes frequencies and
amplitude due to nonlinearities \cite{HOMENJP}.  In these cases,
numerical optimization of the ion's quantum dynamics presents itself
as a well-adapted and efficient approach capable of providing
high-fidelity control solutions.

The results presented in this paper provide us with a
systematic recipe, based on a single parameter (the relative wave
packet size $\xi$), to assess which simulation and control methods are
best suited in different regimes. We observe a crossover between
applicability of the invariant-based IEA, for a very small
wavefunction extension, and that of quantum OCT, when the width of the
wave function becomes comparable with the extension of the potential.
Both methods combined cover the full range of conceivable trap
parameters. That is, no matter what are the trapping parameters,
control solutions for fast, high-fidelity transport are available. In
particular, in the regime $\xi\ll 1$, relevant for ion transport in
chip traps, solutions obtained with the inverse engineering approach
are fully adequate for the 
purpose of achieving high-fidelity quantum operations. This provides a 
major advantage in terms of efficiency over optimization algorithms based
on the solution of the Schrödinger equation. The latter in turn becomes
indispensable when processes involving motional excitations inside the
trap and/or other anharmonic effects are relevant. In this case, the
numerical quantum OCT method demonstrated in this paper provides a
comprehensive way to deal with the manipulation of the ions’ external
states.

\ack
KS, UP, HAF and FSK thank Juan
Gonzalo Muga and Mikel Palmero for the discussions about the invariant
based approach. HAF thanks Henning Kaufmann for useful contributions to the
numerical framework.
The Mainz team acknowledges financial support by the Volkswagen-Stiftung, the
DFG-Forschergruppe (FOR 1493) and the EU-projects DIAMANT (FP7-ICT),
IP-SIQS, the IARPA MQCO project and the MPNS COST Action MP1209. MHG and CPK are grateful to
the DAAD for financial support. SM, FSK and TC acknowledge support from
EU-projects SIQS, DIAMANT and PICC and from the DFG
SFB/TRR21. MHG, SM, TC and CPK enjoyed the hospitality of KITP and
acknowledge support in part by the National Science Foundation under
Grant No. NSF PHY11-25915.

\appendix
\section{Quantum wavepacket propagation with a moving Fourier grid}\label{subs:qmprop}

For transport processes using realistic trap parameters, naive
application of the standard Fourier grid
method~\cite{RonnieReview88,RonnieReview94} will lead to unfeasible
grid sizes. This is due to the
transport distance being usually 3 to 5 orders of magnitude larger
than the spatial width of the wavepacket and possible acceleration of
the wavepacket requiring a sufficiently
dense coordinate space grid. To limit the number of grid points, a
\emph{moving grid} is introduced. Instead of using a spatial grid that
covers the entire transport distance, the grid is defined to only
contain the initial wavepacket, in a window between $x_{\min}$ and
$x_{\max}$. The wavepacket $\Psi(x, t_0)$ is now propagated for a
single time step to
$\Psi(x, t_0+dt)$. For the propagated wave function, the expectation value
\begin{equation}
  \left\langle x \right\rangle
  = \int_{x_{\min}}^{x_{\max}}
   \Psi^{*}(x, t_0 + dt)\, x \, \Psi(x, t_0 + dt) \dd x
\end{equation}
is calculated, and from that an offset is obtained,
\begin{equation}
  \bar{x} = \left\langle x \right\rangle  - \frac{x_{\max} - x_{\min}}{2}\,,
\end{equation}
by which $x_{\min}$ and $x_{\max}$ are shifted.
The wavepacket is now moved to the center of the new grid,
and the propagation continues to the next time step.

The same idea can also be applied to momentum space. After the
propagation step, the expectation value $\left\langle k \right\rangle$
is calculated and stored as an offset $\bar{k}$. The wave function is
then shifted in momentum space by this offset, which is achieved by
multiplying it by $e^{-i\bar{k}x}$. This cancels out the fast
oscillations in $\Psi(x,t_0 + dt)$. When applying the kinetic operator
in the next propagation step, the offset has to be taken into account,
i.e.,
the kinetic operator in momentum space becomes $(k+\bar{k})^2/2m$.

\begin{figure}[tbp]
\begin{center}
\includegraphics{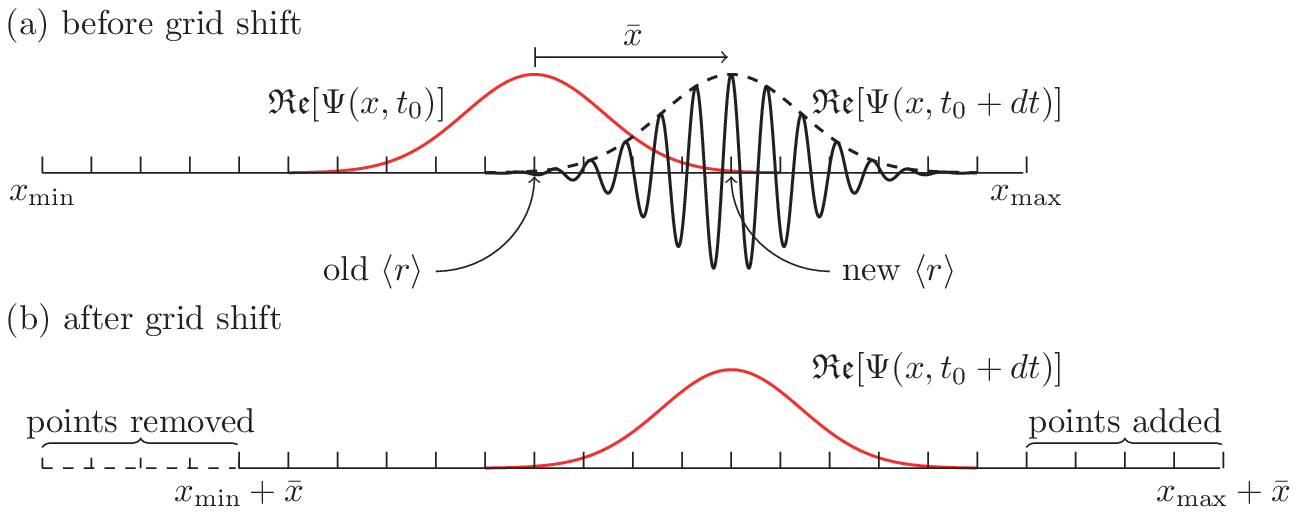}
\caption{Illustration of the moving grid procedure. The propagation of
  the wave function $\Psi(x,t_{0})$ for a single time step is shown in
  (a). The resulting wave function has moved in position and has
  non-zero momentum. After shifting the grid in coordinate and
  momentum space, the propagated wave function is now centered on the
  new grid and has zero momentum (b).}
\label{fig:moving_grid}
\end{center}
\end{figure}

The combination of the moving grid in coordinate and momentum space
allows to choose the grid window with the mere requirement of being
larger than the extension of the wavepacket at any point of the
propagation.  We find typically 100 grid points to be sufficient
to represent the acceleration within a single time step. The procedure
is illustrated in Fig.~\ref{fig:moving_grid} and the steps of the
algorithm are summarized in Table~\ref{tab:howtomovgrid}.
\begin{table}[tbp]
  \caption{\label{tab:howtomovgrid}Necessary steps for wavepacket propagation over long distances.}
\centering
\begin{tabular}
{l  l  l  l }
\br
& & Mathematical step& Possible implementation \\
\mr
  1. & Calculate position mean & $\langle x\rangle=\bra{\Psi}\Op{x}\ket{\Psi}$ & $\langle x\rangle=\sum_i x_i \Psi_i^*\Psi_i$ \\
  2. & Transform to momentum space &  & $\{\Phi_i\}=\mathcal{FFT}(\{\Psi_i\})$ \\
     3. & Calculate momentum mean & $\langle p\rangle=\bra{\Psi}\Op{p}\ket{\Psi}$ & $\langle p\rangle=\sum_i \hbar k_i \Phi_i^*\Phi_i$ \\
  4. & Shift position & $\ket{\Psi}\rightarrow\exp\left(\frac{i}{\hbar}\langle x\rangle \Op{p}\right)\ket{\Psi}$ & $\Phi_i\rightarrow\exp\left(i k_i \langle x\rangle\right)\Phi_i$\\
  5. & Transform to position space &  & $\{\Psi_i\}=\mathcal{FFT}^{-1}(\{\Phi_i\})$ \\
  6. & Shift momentum & $\ket{\Psi}\rightarrow\exp\left(\frac{i}{\hbar}\langle p\rangle \Op{x}\right)\ket{\Psi}$ & $\Psi_i\rightarrow\exp\left(\frac{i}{\hbar} \langle p\rangle x_i\right)\Psi_i$\\
7. & Update classical quantities &  & $x_{cl}+=\langle x\rangle, p_{cl}+=\langle p\rangle$\\       \br
\end{tabular}
\end{table}

\section*{References}


\providecommand{\newblock}{}

\end{document}